# Gadolinium scandate by high-pressure sputtering for future generations of high-κ dielectrics

P. C. Feijoo[1], M. A. Pampillon[1], E. San Andres´ [1] and J. L. G. Fierro[2]

1 Departamento Física Aplicada III: Electricidad y Electrónica, Facultad de CC. Físicas, Universidad Complutense de Madrid, Av. Complutense S/N, Madrid E-28040, Spain

2 Instituto de Catálisis y Petroleoquímica, Consejo Superior de Investigaciones Científicas, C/Marie Curie 2, Cantoblanco E-28049, Spain

**Abstract**

Gd-rich gadolinium scandate ($Gd_{2-x}Sc_xO_3$) was deposited by high-pressure sputtering on (100) silicon by alternating the deposition of <0.5 nm thick films of its binary components: $Sc_2O_3$ and $Gd_2O_3$. The formation of the ternary oxide was observed after the thermal treatments, with a high increase in the effective permittivity of the dielectric (up to 21). The silicon diffuses into the $Gd_{2-x}Sc_xO_3$ films, which show an amorphous character. After the annealing no interfacial silicon oxide is present. $C_{HF}$–$V_G$ curves indicated low hysteresis (55 mV) and a density of interfacial defects of $6 \times 10^{11}$ eV$^{-1}$ cm$^{-2}$.

## 1. Introduction

The downscaling of metal–oxide–semiconductor field-effect transistors (MOSFETs) led in 2008 to the replacement of the gate dielectric SiON for high- permittivity (high κ) dielectrics based on hafnium [1, 2]. This permitted the continuation of the equivalent oxide thickness (EOT) evolution with no increase in the leakage density current. However, future scalability of Hf-based dielectrics is compromised by the low crystallization temperature of the $HfO_2$ [3]. Gadolinium scandate ($GdScO_3$) already emerged as a candidate to replace the SiON gate dielectric, due to its high dielectric constant, energy band gap and thermal stability [4, 5, 6]. Haeni *et al* reported a relative dielectric constant of 20–30 for single crystals depending on the lattice direction [7]; several works determined an optical band gap larger than 5 eV, even for thin films [8, 9, 10]. Zhao *et al* investigated the thermal stability of GdScO$_3$ [11] and they found that GdScO$_3$ films remain amorphous up to 1000 °C. Its suitable properties have made this ternary rare earth oxide arise again as a competitor for the third generation of high-κ materials.

So far, GdScO$_3$ thin films have been deposited on silicon by off-axial pulsed layer deposition [11], e-beam evaporation [12], atomic layer deposition [13] and liquid injection metalorganic chemical vapor deposition [14], demonstrating that amorphous GdScO$_3$ films possess a dielectric constant of 22–23. Roeckerath *et al* fabricated the first MOSFET with the GdScO$_3$ gate oxide on silicon-on-insulator (SOI) substrates, using a gate last process [15, 16]. They reported a density of interface defects $D_{it}$ of $3 \times 10^{11}$ eV$^{-1}$ cm$^{-2}$ and a capacitance ($C_{HF}$–$V_G$) hysteresis of only 50 mV. The leakage current densities remained below $3 \times 10^{-10}$ A cm$^{-2}$ at 1 V for a 2.2 nm capacitance equivalent thickness.

This work demonstrates the gadolinium scandate (Gd$_{2-x}$Sc$_x$O$_3$) growth by high- pressure sputtering (HPS). In this nonconventional sputtering system, the working pressure is around 1 mbar, three orders of magnitude higher than in typical systems. With this pressure, the mean free path of the sputtered particles is around 500 μm, much lower than the target-to-substrate distance (2.5 cm). The particles thus reach the substrate with low energy by a pure diffusion process, which prevents the damage of the substrate and the growing film.

Our procedure to grow Gd$_{2-x}$Sc$_x$O$_3$ was depositing the nano-laminates of scandium oxide and gadolinium oxide (Sc$_2$O$_3$ and Gd$_2$O$_3$). Our HPS system is equipped with a programmable mechanized arm that controls the target positions, so we can switch between materials without breaking the vacuum. After deposition, the nanolaminates were annealed in *forming gas* (FGA) to promote intermixing. This way, we could control the composition of Gd$_2$−$x$Sc$_x$O$_3$ through the thicknesses of the layers of the binary oxides. We evaluated the physical properties of the films by x-ray photoemission spectroscopy (XPS), Fourier transform infrared spectroscopy (FTIR) and transmission electron microscopy (TEM). Platinum gated metal–insulator– semiconductor (MIS) devices were fabricated to determine the leakage current, the high-frequency capacitance and conductance ($C_{HF}$–$V_G$ and $G$–$V_G$) curves, the $C_{HF}$ hysteresis and density of interface defects $D_{it}$. Also, we compare the physical and electrical properties of $Gd_{2-x}Sc_xO_3$ to those of its binary constituents.

2. **Experiment**

Two kinds of substrates were used for the high-κ thin films deposition: high-resistivity and low-resistivity Si wafers. In the first place, 2 inch double side polished (100) n-Si wafers with a resistivity of 200–1000 cm were used for the physical characterization of the deposited films. Secondly, high-κ thin films were grown on 2 inch single side polished (100) n-Si wafers, with a resistivity of 1.5–5.0 cm for the fabrication of Pt gated MIS capacitors.

To design the deposition processes of $Gd_{2-x}Sc_xO_3$, we chose conditions according to previous works on $Sc_2O_3$ and $Gd_2O_3$ [17, 18]. There, we concluded that high pressures produce high-κ films with lower densities of interfacial defects and lower capacitance hysteresis for both binary materials. Thus, thin layers were grown at a pressure of 1.0 mbar (higher pressures could have prompted instabilities in the plasma, so that the repeatability of the processes would not be ensured). A radio frequency (rf) power of 40 W was used. Table 1 summarizes the results obtained for the HPS deposition of $Sc_2O_3$ and $Gd_2O_3$ at 1.0 mbar. The mean growth rates of $Sc_2O_3$ and $Gd_2O_3$ are 0.22 and 0.33 nm min$^{-1}$ respectively. Using this estimation, we tried to control the stoichiometry of the $Gd_{2-x}Sc_xO_3$ layers. We defined the processes to deposit ∼6 nm of $Sc_2O_3$, $Gd_2O_3$ and $Gd_{2-x}Sc_xO_3$ from sputtering of the $Sc_2O_3$ and $Gd_2O_3$ targets. Substrate temperature was maintained at 200 °C during growth. The ternary oxides were grown by alternating the layers of $Sc_2O_3$ and $Gd_2O_3$ with thicknesses of less than 0.5 nm. The cycles were repeated from 8 to 12 times.

We pointed to Gd-rich $Gd_{2-x}Sc_xO_3$ because of two reasons: first, $Gd_2O_3$ presented a thinner $SiO_x$ interface than $Sc_2O_3$ and thus it possesses a higher stability with Si; secondly, according to [19], Gd-rich $Gd_{2-x}Sc_xO_3$ have a higher relative permittivity than other compositions (with a value slightly above 25). In addition, [20] concludes that κ values above 20 can be achieved with different cation (Gd or Sc) compositions.

During the whole deposition time, rf power was applied to both targets simultaneously by using a separate rf source for each cathode. This way, even for the depositions of the binary oxides, plasma was sputtering both targets. In order to grow the binary materials, only one of the targets was placed over the substrate. In the depositions of the binary oxides, only one rf source would have sufficed, but we wanted to check if there was contamination from the other target during growth.

MIS capacitors were fabricated according to the following process flow. First, a 200 nm thick field oxide ($SiO_2$) was thermally grown on the Si substrates to insulate the devices and build up pads (the device contact). Square windows with sizes from 10 × 10 to 700 × 700 $\mu m^2$ were opened on the $SiO_2$ using a positive photoresist process and immersing the samples in a buffered HF solution. After photoresist removal, samples were cleaned by a standard RCA (*Radio Corporation of America*) process. The wafers were then submerged in a 1:50 HF solution for 30 s to remove the native $SiO_2$ within the opened windows just before the introduction to the HPS chamber. After high-κ dielectric deposition by HPS, the contacts and their pads were defined by a negative photoresist process. Top electrode metals were deposited by e-beam evaporation and lifted off. Finally, a 100 nm Ti/200 nm Al stack was evaporated on the entire back surface of the samples to form the bulk contact.

A 10 nm Pt/70 nm Al stack was used as top electrode in the samples with $Sc_2O_3$ and $Gd_2O_3$ as gate dielectrics. The Al capping and the thin Pt layer were motivated by adhesion problems of Pt on the high-κ dielectrics. After the optimization of the second lithography step, we evaporated successfully 50 nm Pt layers on the $Gd_{2-x}Sc_xO_3$ films. The noble metal Pt does not react with the high-κ material so we could measure the plain properties of three insulators: $Sc_2O_3$, $Gd_2O_3$ and Gd-rich $Gd_{2-x}Sc_xO_3$.

The bonding structure of the dielectric films was analyzed by FTIR spectroscopy using a Nicolet Mangna-IR 750 Series II in transmission mode at normal incidence. Spectra were substrate corrected. XPS spectra were obtained by a VG Escalab 200R spectrometer equipped with a MgK$_\alpha$ x-ray source ($hv$ = 1254.6 eV), powered at 120 W.

We measured high-frequency (100 kHz) $C_{HF}$–$V_G$ and $G$–$V_G$ curves of the MIS devices by an Agilent 4294A impedance analyzer and the leakage current density ($J_G$–$V_G$) at ambient temperature by a Keithley 2636A system. Then, the samples were exposed to a 20 min FGA at 300 °C and the electrical characterization was repeated. Finally, the characterization was made for the third time after a second FGA at 450 °C.

We extracted parameters such as the EOT, the flatband voltage $V_{FB}$ and the $D_{it}$. This way, we analyzed the influence of the FGA on the dielectric permittivity and the electrical performance. TEM images were taken from the Pt/$Gd_{2-x}Sc_xO_3$/Si MIS devices after the annealings with a Tecnai T20 microscope from FEI at 200 keV. TEM samples were prepared by a DualBeam focused ion beam system from FEI.

### 3. Results and discussion

*3.1. Structural characterization*

We designed deposition processes to obtain $Gd_{2-x}Sc_xO_3$ films with different compositions. The most relevant results were obtained for the samples with a surface composition of

$Gd_{1.8}Sc_{0.2}O_3$, for which the subsequent discussion is centered. This stoichiometry was measured by XPS, using the Sc 2p and Gd $3d_{5/2}$ peaks (figure 1(*a*)). Additionally, the XPS composition measurements from figure 1(*b*) ensure that when a binary $Sc_2O_3$ or $Gd_2O_3$ film is grown, it presents no trace of the other material, although both targets were being sputtered during deposition.

Figure 2 represents the baseline corrected FTIR spectra of the thin films deposited on Si. They all show a peak at 669 cm$^{-1}$ that corresponds with the C–O bond from the $CO_2$ in the chamber during FTIR measurements. The band at 1025 cm$^{-1}$ indicates that all three materials have a thin $SiO_x$ at the dielectric/Si interface. From the area of the band, we can conclude that the $Sc_2O_3$ produces a thicker $SiO_x$ interface layer than that of $Gd_2O_3$. The area of this band is represented forthethreehigh-κ materialsintheinsetoffigure2.Regarding the $Gd_{2-x}Sc_xO_3$ film, it presents an $SiO_x$ film comparable with the $Gd_2O_3$ interface. Then, $Gd_{2-x}Sc_xO_3$ is at least as stable as $Gd_2O_3$ in contact with Si. The shift of the maximum toward lower wavenumbers indicates a sub-stoichiometric or stressed $SiO_x$ for the three high-κ materials [21].

Figure 3 presents a TEM image of an MIS device that consists of a $Pt/Gd_{2-x}Sc_xO_3/Si$ stack after the two FGAs at 300 and 450 °C. This image shows an amorphous dielectric with a thickness of 7 ± 1 nm. Below the dielectric, the image shows the typical diffraction patterns of the mono-crystalline Si, while above the dielectric, a poly-crystalline Pt is found.

The $Pt/Gd_{2-x}Sc_xO_3$ interface is abrupt, as expected, due to the low reactivity of Pt. No clear interfacial $SiO_x$ layer between the $Gd_{2-x}Sc_xO_3$ and the Si substrate is distinguished. However, we can distinguish two different intensities in the $Gd_{2-x}Sc_xO_3$ film that suggest a reaction between the dielectric layer and the thin interface $SiO_x$. The TEM image suggests a silicate formation with a thickness of around 4 nm. This reaction may take place during the annealings and can comprise the effective permittivity of the dielectric, since silicates generally present lower dielectric constants than the oxides.

Figure 4 shows an in-depth sweep in the intensity of energy-dispersive x-ray signals, taken during TEM images acquisition. In order to obtain this graph, an electron beam swept the $Pt/Gd_{2-x}Sc_xO_3/Si$ stack and an x-ray detector registered the intensity of a peak of each atom as a function of the position of the beam. This allows the identification of the atoms along the stack. This graph confirms again the presence of $Gd_{2-x}Sc_xO_3$ as the gate dielectric, sandwiched between Pt and Si. The Si signal enters in the $Gd_{2-x}Sc_xO_3$ region, supporting the idea of the reaction between the high-κ dielectric and the interface $SiO_x$ (the silicate formation).

*3.2. Electrical characterization*

The $C_{HF}$–$V_G$ and $G$–$V_G$ characteristics of an MIS device with a $Gd_{2-x}Sc_xO_3$ dielectric unannealed and after the FGAs at 300 and 450 °C are shown in figures 5(*a*) and (*b*), respectively. The $C_{HF}$ curves indicate that the capacitance measured in accumulation, $C_{ins}$, increases with the FGA. This means that the dielectric and the substrate do not react to form $SiO_x$, and thus $Gd_{2-x}Sc_xO_3$ is stable in contact with Si. Furthermore, the effective permittivity of the dielectric is increasing, which suggests that during the FGA $Gd_{2-x}Sc_xO_3$ is being formed from its binary constituents. The shifts in the $V_{FB}$ indicate that the charge in the dielectric is changing with the FGA.

The peak in the conductance (figure 5(*b*)) is due to the density of interfacial defects $D_{it}$. Although to obtain quantitative results we must remove the influence of the series resistance, qualitatively we can observe a decrease in the conductance peak with annealing temperature, which means a lower $D_{it}$ after the FGA and a higher quality of the interface. This is caused by the passivation of the dangling bonds in the dielectric/Si interface with H atoms. We will quantitatively confirm this result later with the analysis of the calculated $D_{it}$.

In summary, the capacitance of the Pt/$Gd_{2-x}Sc_xO_3$/Si stack increases after the annealings, and thus the FGA reduces the EOT from 3.0 to 1.3 nm. As we observe in the TEM image in figure 3, the $SiO_x$ does not grow at the dielectric/Si interface. The effective relative permittivity of the dielectric layer increases with the annealing, indicating the formation of Gd–Sc bonds. For this film, we can calculate a relative dielectric constant of ∼21, using the measured thickness of 7 nm. Kittl *et al* reported a similar value for Gd-rich $Gd_{2-x}Sc_xO_3$ [19]. This relative permittivity lies within the range of 10–30 that is necessary in the future generations of Si MOSFETs [22], and it is close to the value of 25 that would minimize the short channel effect in transistors (due to the fringing field in source and drain) [23]. This permittivity confirms $GdScO_3$ as a strong candidate to replace the Hf silicates, whose permittivity is much lower (for $HfSiO_4$, the relative permittivity is around 11) [24]. Besides, the interface traps are partially passivated by the H atoms during the FGA, improving the quality of the interface. In the following paragraphs, we compare the electric behavior of the ternary oxide with its binary constituents, $Gd_2O_3$ and $Sc_2O_3$.

Figure 6 illustrates the evolution of the EOT with the FGA for the Pt gated MIS devices with binary oxides. As we have just pointed out, the EOT of $Gd_{2-x}Sc_xO_3$ dramatically decreases with FGA temperature, due to the formation of the ternary from the layers of the binary oxides and the increment of the permittivity. On the other hand, the EOT increases 0.5 nm for $Gd_2O_3$ and 1 nm for the $Sc_2O_3$. This means that a layer of $SiO_x$ is growing at the dielectric/Si interface for both materials, although $Gd_2O_3$ is slightly more stable than $Sc_2O_3$ in contact with Si. The behavior of the binary oxides is completely opposed to the trend that follows the ternary oxide. This highlights the higher permittivity of $Gd_{2-x}Sc_xO_3$ as compared to the binary oxides—in our previous works, for the same deposition system and these deposition conditions, the scandium oxide presented a relative permittivity of 9 [25], while the gadolinium oxide, a relative permittivity of 11 [18].

The $D_{it}$ values calculated with the conductance method are represented in figure 7 for the unannealed MIS devices and after the FGAs. The passivation of the dangling bonds in the interface reduces $D_{it}$ for the three high-κ materials. For $Gd_{2-x}Sc_xO_3$, the passivation effect seems to saturate at 300 °C, improving the $D_{it}$ one order of magnitude compared to the unannealed case and reaching values of $6 \times 10^{11}$ eV$^{-1}$ cm$^{-2}$. However, $D_{it}$ is hard to calculate after the second annealing due to the high conductance in accumulation. The saturation also appears for the $Gd_2O_3$ devices, for which an even lower density of defects is obtained. $Sc_2O_3$ presents an even better interface and the saturation is not observed. The better interface quality of the binary oxides can be related to the regrowth of the $SiO_x$ in the dielectric/Si interface. In fact, a thicker and relaxed $SiO_x$ would also explain the lower $D_{it}$ in the case of the $Sc_2O_3$.

Table 2 presents the flatband voltage of the MIS capacitors extracted from the $C_{HF}$–$V_G$ curves (starting from inversion) and the associated dielectric charge density $Q_{ss}$. We calculated the dielectric charge according to equation (1). We have assumed a work-function difference $\phi_{ms}$ of 1.3 eV between Pt (workfunction of 5.6 eV) and Si (a resistivity of 1.5–5.0 cm means a work-function of around 4.3 eV):

$$V_{FB} = \phi_{ms} - \frac{Q_{ss}}{C_{ins}} \quad (1)$$

The devices fabricated with the three dielectrics possess a $V_{FB}$ of 0.2–0.4 V before the annealings, which implies a dielectric charge in the range of $6 \times 10^{12}$ cm$^{-2}$. The FGAs increase the dielectric charge up to around $10^{13}$ cm$^{-2}$ for the ternary oxide and both binary oxides. This decreases significantly the flatband voltage in all cases, which reaches negative values. This positive charge can be related to the hydrogen passivation of the defects in the dielectrics during the FGAs.

The $C_{HF}$ hysteresis curves of the MIS devices were measured after the annealings. The results are also presented in table 2. A flatband voltage shift $V_{FB}$ of only 55 mV is obtained for the Pt/Gd$_{2-x}$Sc$_x$O$_3$/Si devices, as can be observed in figure 8. This corresponds to a density of slow states of $10^{11}$ cm$^{-2}$, below the density of slow defects of the binary oxides (around $5 \times 10^{11}$ cm$^{-2}$). This shift is very small compared to the next generation requirements [26, 27], which is another advantage of the ternary rare earth oxide. This low hysteresis must be due to the passivation of the slow response defects during the FGA and a low density of defects in the bulk dielectric.

Finally, we discuss the leakage current of the MIS devices with the different high-κ dielectrics. Figure 9(*a*) shows the $J_G$–$V_G$ curves for the gadolinium scandate before and after the annealings. $J_G$ clearly increases one order of magnitude after the annealing at 300°C and about five orders of magnitude after the annealing at 450 °C. The increase in the leakage current can be caused by a densification of the dielectric film, i.e. a reduction of its thickness, with an increase of the tunneling current. However, a dielectric thickness of 7 nm should be too high to observe any tunneling. A possible origin could also be the creation of conduction paths, but we have seen that both the defects in the interface and the slow defects are passivated by the annealing. The most feasible cause is then the dissolution of the interfacial SiO$_x$ into the Gd$_{2-x}$Sc$_x$O$_3$ film producing a silicate interface with a lower electron barrier. This would increase the number of electrons that can flow from the Si conduction band into the insulator.

In figure 9(*b*), we compare the leakage current density at 1 V for the three dielectrics. The unannealed scandium oxide presents a higher leakage current than the other two materials, but it remains at almost the same value after the FGAs. This can be due to the thick SiO$_x$ interface between the Sc$_2$O$_3$ and the Si and its regrowth with the FGAs, which provides a thick and high barrier for the electrons. On the other hand, the gadolinium oxide leakage current, such as $Gd_{2-x}Sc_xO_3$, increases with the FGA. However the values are lower, probably due to a slightly thicker dielectric, together with a thicker $SiO_x$ interface. In the inset of figure 9(*b*), the $J_G$ is represented as a function of the EOT for the MIS devices with the gadolinium scandate gate dielectric. The leakage current reaches $10^{-1}$ A cm$^{-2}$ for 1.3 nm EOT at 1 V.

## 4. Summary and conclusions

In this work, we have deposited gadolinium scandate by HPS from the sputtering of $Sc_2O_3$ and $Gd_2O_3$ targets, alternating sub-nanometric thick layers. We explore the electrical and physical characterization of the films, with a thorough comparison with films of the binary oxides.

The composition of the Gd-rich $Gd_{2-x}Sc_xO_3$ was measured by XPS. The most relevant results were obtained from the $Gd_{1.8}Sc_{0.2}O_3$ surface composition. The FTIR spectra revealed a good stability of the nano-laminates of $Gd_2O_3$ and $Sc_2O_3$ after the deposition on Si. MIS capacitors were fabricated to assess the electrical performance of the dielectric. High frequency capacitance measurements showed that FGAs increase the EOT in the case of the binary oxides but in the case of the ternary, the EOT greatly decreases, reaching values of 1.3 nm. This result points to the formation of $Gd_{2-x}Sc_xO_3$ from the binary components due to the annealings. The TEM images after the annealing at 450 °C displayed a 7 nm amorphous $Gd_{2-x}Sc_xO_3$ layer, with the formation of a 4 nm silicate in the region close to the high-κ dielectric/Si interface. The FGAs also passivated the dielectric/Si interfaces, with a density of defects around $6 \times 10^{11}$ eV$^{-1}$ cm$^{-2}$. The Pt/$Gd_{2-x}Sc_xO_3$/Si MIS capacitors also present a small hysteresis after the FGAs and low leakage currents. Finally, we attained an effective relative permittivity of 21, which is a promising value for the application of this material in future generations of MOSFET devices.

**Acknowledgments**

The authors would like to acknowledge the collaboration of the CAI de Tecnicas Físicas and the CAI de Espectroscopía y Espectrometría of the Universidad Complutense de Madrid.

This work was financed by the Spanish MINECO under project TEC2010-18051. The work of MAP and PCF was funded by the FPI program (BES-2011-043798) and FPU program (AP2007-01157), respectively.

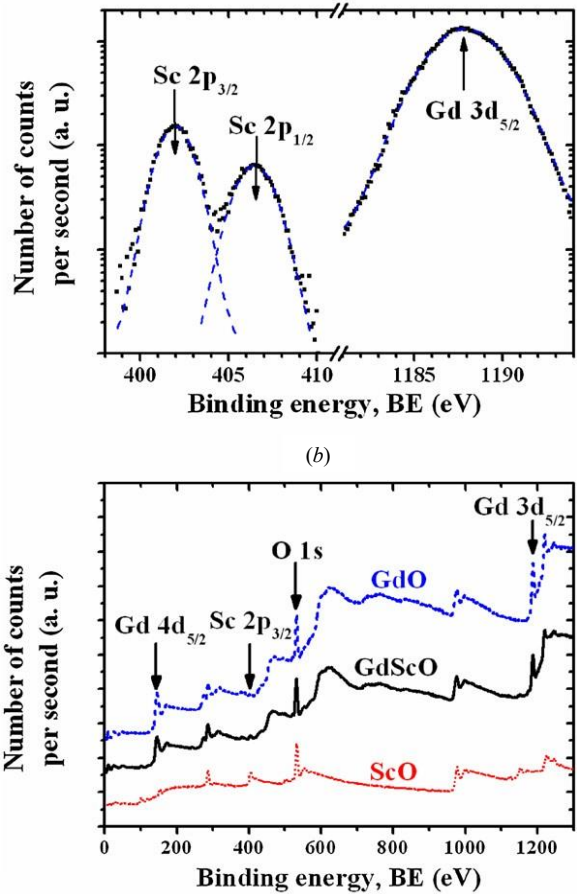

**Figure 1.** (*a*) Sc 2p and Gd $3d_{5/2}$ XPS peaks of the $Gd_{1.8}Sc_{0.2}O_3$ films. From peak areas and sensibility factors (1.6 for Sc 2p and 3.41 for Gd $3d_{5/2}$), we calculated the composition.

(*b*) XPS spectra of the $Sc_2O_3$, $Gd_{1.8}Sc_{0.2}O_3$ and $Gd_2O_3$. $Sc_2O_3$ spectrum presents no trace of Gd peaks although both targets were sputtered by plasma at the same time during deposition. The same happens for $Gd_2O_3$.

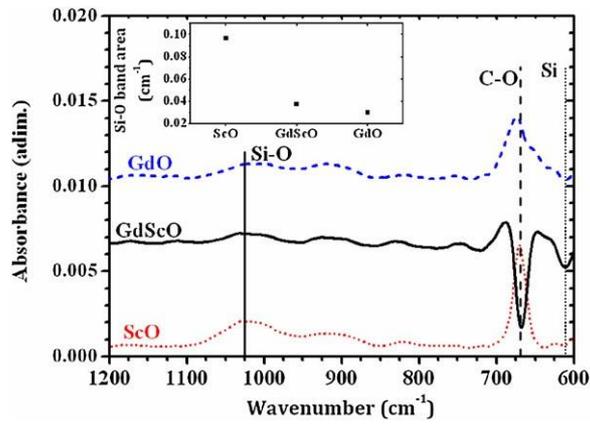

**Figure 2.** FTIR spectra of the $Sc_2O_3$, $Gd_{2-x}Sc_xO_3$ and $Gd_2O_3$ films. We corrected substrate absorbance and baseline, and the spectra were shifted vertically for clarity. The inset shows the Si–O band areas for the three films.

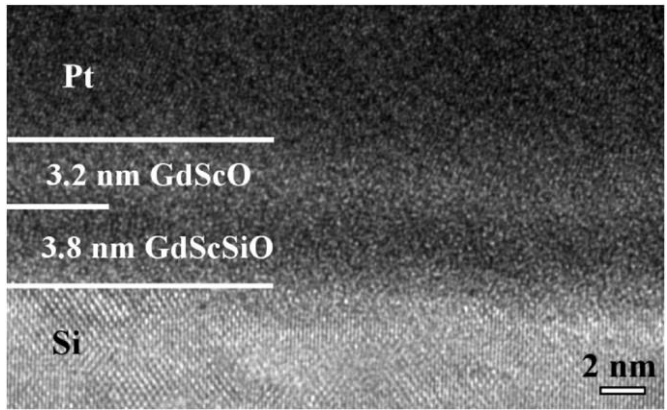

**Figure 3.** TEM image of the Pt/Gd$_{2-x}$Sc$_x$O$_3$/Si MIS capacitor, after the FGAs at 300 and 450 °C. The darker region in the dielectric film indicates the formation of a silicate.

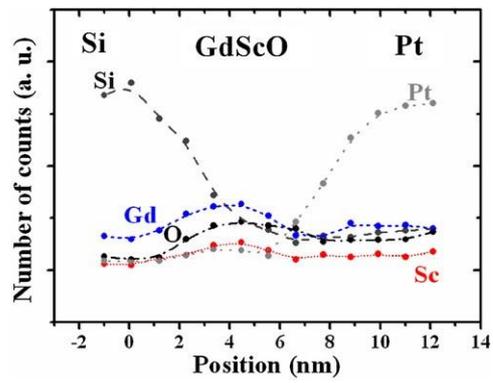

**Figure 4.** Intensity of electron excited x-rays as a function of position in a Pt/Gd$_{2-x}$Sc$_x$O$_3$/Si MIS capacitor. Dashed and pointed lines are drawn as a guide to the eye.

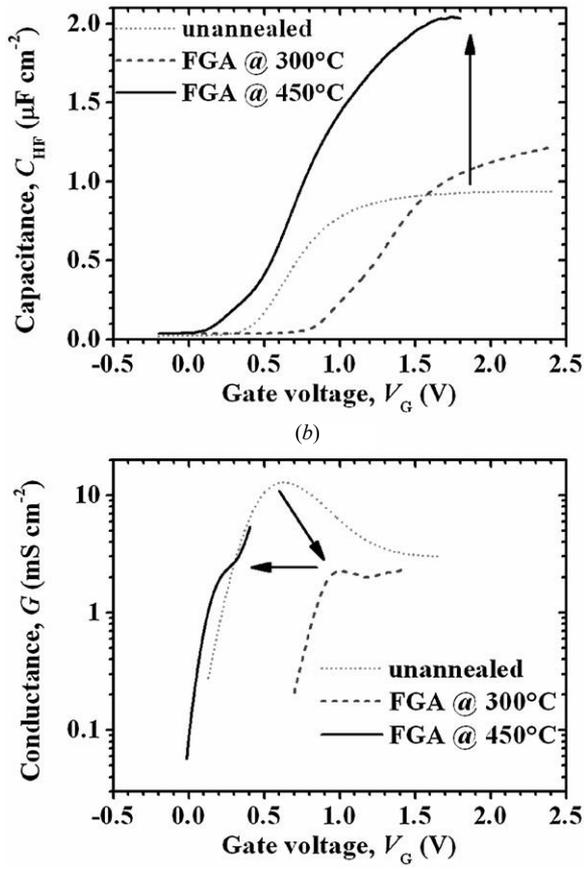

**Figure 5.** (*a*) $C_{HF}$–$V_G$ and (*b*) $G$–$V_G$ characteristics of the Pt/Gd$_{2-x}$Sc$_x$O$_3$/Si device, after device fabrication, after the FGA at 300 °C and after the FGA at 450 °C.

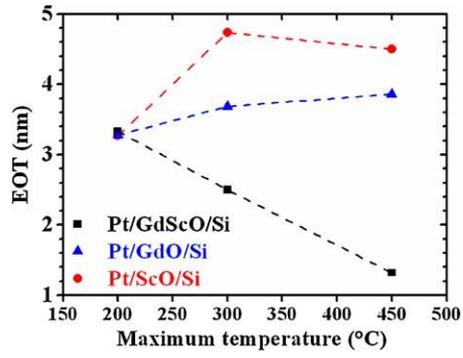

**Figure 6.** EOT as a function of the maximum processing temperature for the Sc$_2$O$_3$, Gd$_2$O$_3$ and Gd$_{2-x}$Sc$_x$O$_3$ dielectrics. Dashed lines are drawn to guide the eye.

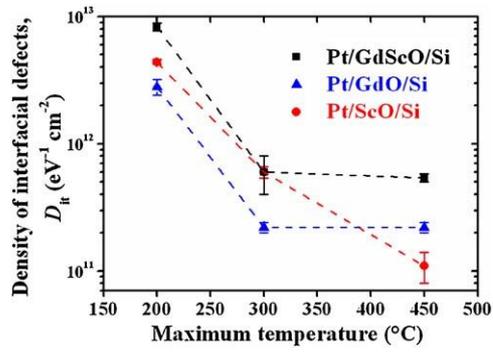

**Figure 7.** $D_{it}$ as a function of the maximum processing temperature for the $Sc_2O_3$, $Gd_2O_3$ and $Gd_{2-x}Sc_xO_3$ dielectrics. Dashed lines are drawn as a guide to the eye.

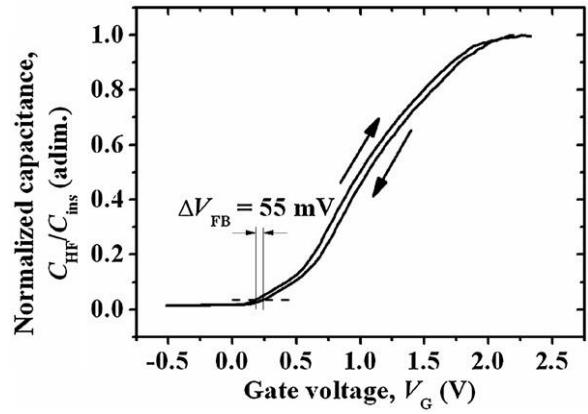

**Figure 8.** $C_{HF}$–$V_G$ hysteresis curve for the Gd$_{2-x}$Sc$_x$O$_3$ sample. It presents a flatband voltage shift of 55 mV.

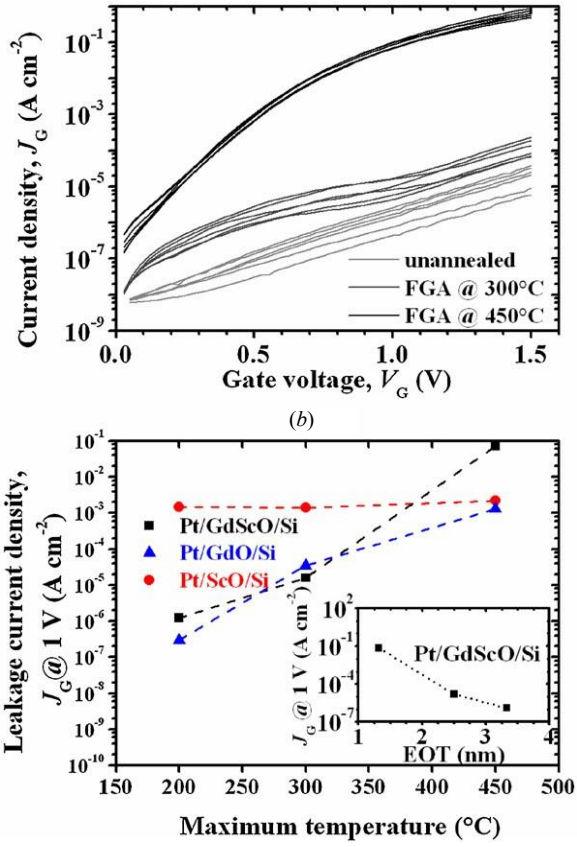

**Figure 9.** (*a*) $J_G$–$V_G$ curves of the $Gd_{2-x}Sc_xO_3$ dielectric before and after the FGA. (*b*) Leakage current density at 1 V for the $Sc_2O_3$,

**Table 1.** Growth parameters of the 1.0 mbar $Sc_2O_3$ and $Gd_2O_3$ deposited films.

| Target | Deposition time (min) | Pressure (mbar) | rf power (W) | Thickness (nm) | Refractive index |
|---|---|---|---|---|---|
| Sc2O3 | 30 | 1.0 | 40 | 6.6 ± 0.2 | 1.440 |
| Gd2O3 | 30 | 1.0 | 40 | 6.6 ± 0.2 | 1.905 |

**Table 2.** Flatband voltage, dielectric charge, flatband voltage shift and charge variance during $C_{HF}$–$V_G$ hysteresis of the MIS capacitors before and after the annealings for the three dielectric materials.

|  | Scandium oxide | | Gadolinium oxide | | Gadolinium scandate | |
| --- | --- | --- | --- | --- | --- | --- |
|  | Unannealed | After annealing | Unannealed | After annealing | Unannealed | After annealing |
| $V_{FB}$ (V) | 0.3 | -0.2 | 0.2 | -0.5 | 0.4 | -0.2 |
| $Q_{ss}$ (cm$^{-2}$) | 6e12 | 7e12 | 6e12 | 1e13 | 6e12 | 1e13 |
| $\Delta V_{FB}$ (mV) |  | 84 |  | 26 |  | 55 |
| $\Delta Q_{ss}$ (cm$^{-2}$) |  | 4e11 |  | 6e11 |  | 1e11 |